\documentclass[aps,prl,twocolumn,letterpaper,superscriptaddress, showpacs, amsmath, amssymb]{revtex4}

\usepackage{graphicx}
\usepackage{amsmath,amssymb}

\begin{document}

\title{Neutron-Proton-Converter Acceleration Mechanism \\ at Subphotospheres of Relativistic Outflows} 

\author{Kazumi Kashiyama} 
\affiliation{Department of Astronomy \& Astrophysics; Department of Physics; Center for Particle \& Gravitational Astrophysics; Pennsylvania State University, University Park, PA 16802}

\author{Kohta Murase} 
\affiliation{Hubble Fellow --- Institute for Advanced Study, Princeton, NJ 08540}
\affiliation{Center for Cosmology and AstroParticle Physics; Department of Physics, The Ohio State University, Columbus, OH 43210}

\author{Peter M$\rm \acute{e}$sz${\rm \acute{a}}$ros}
\affiliation{Department of Astronomy \& Astrophysics; Department of Physics; Center for Particle \& Gravitational Astrophysics; Pennsylvania State University, University Park, PA 16802}

\date{\today}

\begin{abstract}
We study a type of particle acceleration that operates via neutron-proton conversion in inelastic nuclear collisions.  
This mechanism can be expected for relativistic shocks at subphotospheres if relativistic outflows contain neutrons.  
Using a test-particle approximation, we numerically calculate the energy spectrum and the efficiency of accelerated particles, and show that a good energy fraction of the nucleons can be accelerated.  
This mechanism may especially be relevant if the shock is radiation-mediated, and it would enhance the detectability of GeV-TeV neutrinos. 
\end{abstract}

\pacs{98.70.Sa, 95.85.Ry 98.70.Rz \vspace{-0.3cm}}

\maketitle

Cosmological gamma-ray bursts (GRBs) accompany relativistic jets, where luminous prompt $\gamma$ rays are generated. 
Although its emission mechanism has been a long-standing problem (e.g., \cite{Gehrels_Meszaros_2012}), 
internal shocks in unsteady jets have been thought to play an important role~\cite{Rees_Meszaros_1994}.  
In particular, if such shocks happen beyond the photosphere, at which the Thompson optical depth $\tau_{\rm T}\sim1$, 
the prompt $\gamma$ rays may be explained by optically-thin synchrotron emission from electrons accelerated at collisionless shocks~\cite{Rees_Meszaros_1994}, 
although this classical scenario has several difficulties e.g., the radiation efficiency and the inconsistency with the observed spectra~\cite{Zhang_et_al_2007,Preece_et_al_2000}.  

The dissipative photospheric scenario~\cite{Thompson_1994,Meszaros_Rees_2000b} could overcome the above shortages.  
In this scenario, the prompt $\gamma$ rays are attributed to modified thermal~\cite{Thompson_1994,Meszaros_Rees_2000b,2007ApJ...670L..77I,Beloborodov_2010,2013MNRAS.428.2430L} 
and/or synchrotron emissions~\cite{2011ApJ...738...77V,2012ApJ...746..164M} around the photosphere.  
An interesting channel of the subphotospheric dissipation exists in neutron-loaded outflows~\cite{Bahcall_Meszaros_2000,Rossi_et_al_2006,Koers_Giannios_2007}, 
where the hadronuclear reaction between protons and neutrons plays an important role and resulting cascades with Coulomb heating may help to form observed spectra~\cite{Beloborodov_2010}.  
Such a neutron loading is a natural consequence given that the jet is launched from an extremely dense, hot region where the electron capture proceeds~\cite{Beloborodov_2003a}.

In the coming years, neutrino astronomy may provide a breakthrough.  
Assuming that protons are also accelerated at internal shocks, TeV-PeV neutrinos were predicted in the classical scenario~\cite{1997PhRvL..78.2292W}, 
and IceCube~\cite{Ahrens_et_al_2004} has constrained reasonable parameter ranges from the nondetection
\cite{Abbasi_et_al_2012,2012PhRvD..85b7301L,2012ApJ...752...29H,2012PhRvL.108w1101H}. 
Different predictions for the photospheric scenario were also made~\cite{2008PhRvD..78j1302M,2009ApJ...691L..67W,2012ApJ...746..164M}.  
Without invoking non-thermal protons, the inelastic-collision model naturally predicts multi-GeV quasithermal neutrinos~\cite{Bahcall_Meszaros_2000,Meszaros_Rees_2000}, 
which can be detected by the low-energy extension of IceCube, DeepCore~\cite{Abbasi_et_al_2012b,Murase_et_al_2013,2013arXiv1301.4232B}.  

Since the effective area of DeepCore becomes significantly small at lower energies, high-energy neutrinos are crucial in terms of detectability. 
However, deeply under the photosphere ($\tau_{\rm T}\gg1$), there seems to be a theoretical difficulty in proton acceleration, i.e.,  
even if internal shocks occur, the conventional Fermi acceleration~\cite{1981MNRAS.196..135P,Blandford_Eichler_1987, Gallant_Achterberg_1999, Achterberg_et_al_2001,2006MNRAS.366..635L} 
would be inefficient at radiation-mediated shocks~\cite{Levinson_Bromberg_2008,Murase_et_al_2011,Murase_Ioka_2013}.  
Here we demonstrate that the neutron-proton-converter~(NPC) acceleration mechanism, where conversions between neutrons and protons are implemented in the course of the Fermi acceleration, 
can operate in neutron-loaded jets even at the subphotosphere, and a reasonable energy fraction of the neutron flow is transferred to non-thermal nucleons and neutrinos. 

The NPC acceleration was originally proposed by Derishev et al~\cite{Derishev_et_al_2003} with simple analytical considerations. 
However, relativistic-shock accelerations generally depends on details of scattering processes~\cite{2003ApJ...591..954V,2008MNRAS.383.1431A}.
Thus, we here perform Monte-Carlo simulations, which give the resultant spectra of nucleons and neutrinos correctly.   
Also, we for the first time explore the NPC acceleration in relativistic flows at the subphotospheres, where inelastic nuclear collision is the relevant conversion process, 
in the context of the dissipative photospheric model. 

{\it NPC acceleration: A slow slugger.---}
The advantage of invoking NPC acceleration mechanism is highlighted by considering the possible energy gain per acceleration cycle~\cite{Derishev_et_al_2003}. 
As in the the Fermi acceleration, particles which cross the shock experience a Lorentz boost; 
\begin{equation}
\gamma_{\rm d} \rightarrow \gamma_{\rm u} = \Gamma_{\rm rel}\gamma_{\rm d}(1-\beta_{\rm rel}\beta_{\rm d}\mu_{\rm d}),
\end{equation}
for the downstream to the upstream, and 
\begin{equation}
\gamma_{\rm u} \rightarrow \gamma_{\rm d} = \Gamma_{\rm rel}\gamma_{\rm u}(1+\beta_{\rm rel}\beta_{\rm u}\mu_{\rm u}),
\end{equation}
for the upstream to the downstream. 
Here, quantities subscribed u(d) are defined in the up(down)stream rest frame. 
We use $\Gamma_{\rm rel}\approx0.5(\Gamma_{\rm u}/\Gamma_{\rm d}+\Gamma_{\rm d}/\Gamma_{\rm u})$ for the relative Lorentz factor between the shock upstream and the downstream, 
and $\mu$ for the pitch angle relative to the shock normal when crossing the shock.  
In a conversion process, a nucleon loses its energy as 
\begin{equation} 
\gamma \Rightarrow \kappa_{\rm pn}(\gamma -1) +1, 
\end{equation} 
with the inelasticity $\kappa_{\rm pn}$. 
As a result, the energy gain per cycle can be described as
\begin{equation}\label{eq:ene_gain}
\left\langle E_{\rm f}/E_{\rm i} \right\rangle \approx  \kappa_{\rm pn}{}^{N_{\rm co}} \Gamma_{\rm rel}{}^2 (1-\langle \mu_{\rm u} \rangle)(1+ \langle \mu_{\rm d} \rangle),  
\end{equation}
in the relativistic limit ($\gamma\gg1$, $\Gamma_{\rm rel}\gg1$). 
Here $N_{\rm co}$ is the number of inelastic collision in the cycle, and the angled brackets mean the flux ensemble of the particles that cross the shock.  
One can see that $\langle E_{\rm f}/E_{\rm i}\rangle~\propto~\Gamma_{\rm rel}{}^2$ 
unless $\langle\mu_{\rm u}\rangle~\approx~1$; i.e., the particles are isotropized before crossing the shock from the upstream to the downstream. 

In the relativistic-shock acceleration without conversions, $\langle E_{\rm f}/E_{\rm i}\rangle\approx2\Gamma_{\rm rel}{}^2$ can be realized only in the first cycle, 
and $\langle E_{\rm f}/E_{\rm i}\rangle\lesssim2$ in the successive ones~\cite{Gallant_Achterberg_1999,Achterberg_et_al_2001}. 
This is because accelerated protons which cross the shock from the downstream to the upstream are captured by the shock typically in the very early phase of the gyration with 
$\langle\mu_{\rm u}\rangle\approx1-1/\Gamma_{\rm rel}{}^2$.  
In the NPC acceleration, on the other hand, neutrons that cross the shock can go far downstream before being converted to protons.  Then, the converted protons are isotropized in the upstream before being captured by the shock, as long as the gyration frequency is much larger than the conversion frequency (which would be realized in GRB jets). 
In this case, the NPC acceleration provides a larger energy boost per cycle $\langle E_{\rm f}/E_{\rm i}\rangle\approx2\kappa_{\rm pn}{}^{N_{\rm co}}\Gamma_{\rm rel}{}^2$, 
especially for a larger $\Gamma_{\rm rel}$. 
As a tradeoff, the duration of the cycle is essentially determined by the conversion time scale, which makes the NPC acceleration a relatively slow process. 
We note that the NPC acceleration via hadronuclear reactions is ineffective for non-relativistic shocks ($\Gamma_{\rm rel} \approx 1$) due to the inelasticity  
(see \cite{Blasi_et_al_2012} for different converter processes in proton acceleration at non-relativistic shocks). 

\begin{figure}
\includegraphics[width=2.85in]{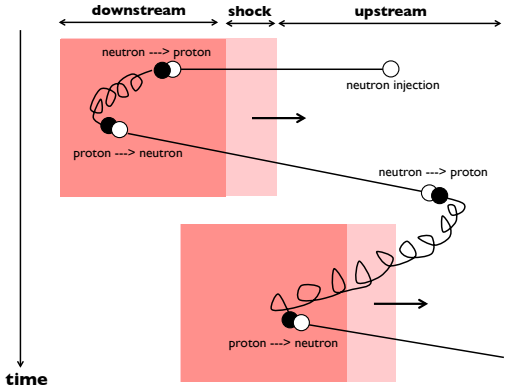}
\caption{Schematic illustration of the NPC cycle.}  
\label{f0}
\vspace{-1.\baselineskip}
\end{figure}

\begin{figure*}
\includegraphics[width=3.25in]{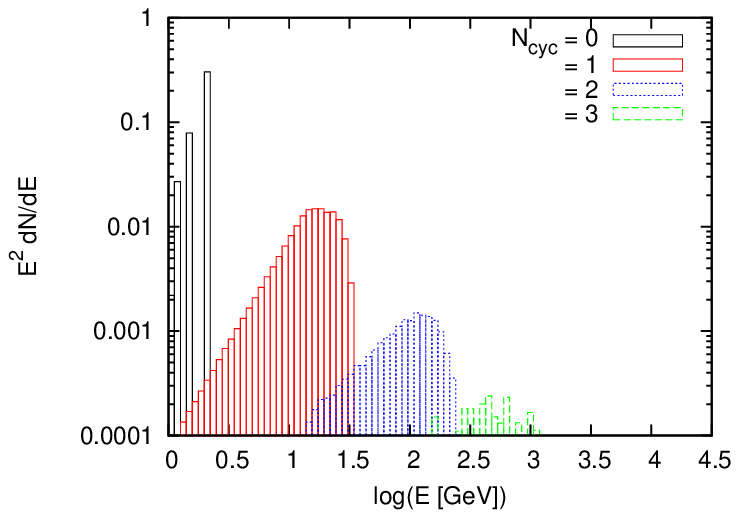}
\includegraphics[width=3.25in]{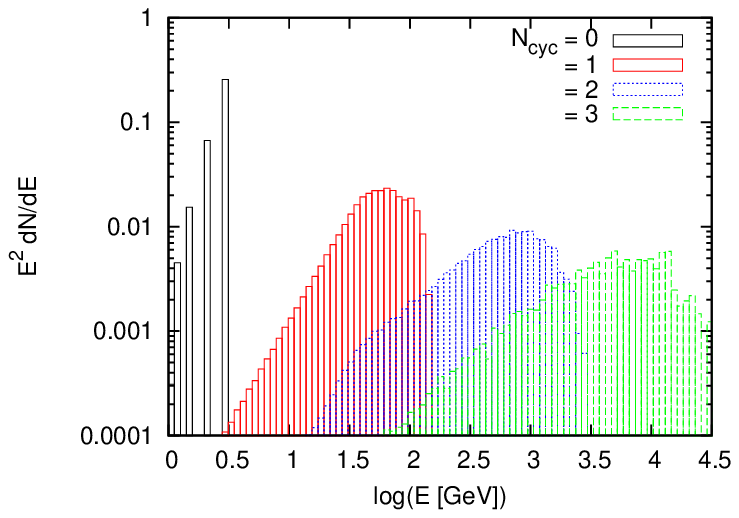}
\caption{The energy spectrum of protons in the downstream for $\Gamma_{\rm rel}=3$ (left) and $5$ (right). 
We set $\gamma_{\rm d,o}=\Gamma_{\rm rel}$, $\tau_{\rm pn}=2$, and $\xi(1)=10^6$. 
The spectra are normalized by the total kinetic energy of the neutron injection.}  
\label{f1}
\vspace{-1.\baselineskip}
\end{figure*}

{\it NPC acceleration in neutron-loaded outflows.---}
Let us consider an internal shock in a neutron-loaded jet at the subphotosphere~$\tau_{\rm T} > 1$. 
A rapid compound flow with $\Gamma_{\rm r}$ and a slow one with $\Gamma_{\rm s}$ collide at $r\approx2\Gamma_{\rm s}{}^2r_{\rm o}\sim2\times10^{11} \ \Gamma_{{\rm s},2}r_{{\rm o}, 7}\ {\rm cm}$, 
where $r_{\rm o}=10^7 \ r_{{\rm o},7}\ \rm cm$ is the launching radius of the jet. 
Hereafter we use $Q_x=Q/10^x$ in CGS units.
The protons would effectively form the shock jump in the sense that length scales of collisionless dissipation or radiation precursor are relatively short.  
One can estimate the Lorentz factor of the shocked region as $\Gamma\approx\sqrt{\Gamma_{\rm r}\Gamma_{\rm s}}\sim600 \ \Gamma_{{\rm r},3.5}{}^{1/2}\Gamma_{{\rm s},2}{}^{1/2}$. 
For a (coasting) slow flow, neutrons are coupled with protons up to the decoupling radius where $\tau_{\rm pn} = n_{\rm u}\sigma_{\rm pn}r/\Gamma_{\rm s} \approx 1$, 
or $r_{\rm dec}\approx L_{\rm iso}\sigma_{\rm pn}/4\pi m_{\rm p}c^3\Gamma_{\rm s}{}^3\sim5\times10^{10} \ L_{\rm iso, 51}\Gamma_{\rm s, 2}^{-3}\ \rm cm$.  
Here $n_{\rm u}\approx L_{\rm iso}/4\pi\Gamma_{\rm s}{}^2r^2m_{\rm p}c^3$ is the baryon number density in the rest frame of the unshocked slow flow~(hereafter upstream), 
and $L_{\rm iso}$ is the isotropic luminosity.   
When internal shocks happen under $r_{\rm dec}$, neutrons are injected into the shocked slow flow~(hereafter downstream) with the relative Lorentz factor
$\Gamma_{\rm rel}\approx0.5(\Gamma/\Gamma_{\rm s}+\Gamma_{\rm s}/\Gamma)\sim3 \ \Gamma_{{\rm r},3.5}{}^{1/2}\Gamma_{{\rm s},2}{}^{-1/2}$. 
For $\tau_{\rm pn}\gtrsim1$, the injected neutrons cause inelastic collisions with a larger length scale~\cite{2013arXiv1303.2612M}, 
producing electrons-positrons, $\gamma$ rays, and neutrinos as the by-products.   
The typical neutrino energy is 
$E^{\rm obs}_{\nu}\approx0.1\Gamma\Gamma_{\rm rel} \tau_{\rm pn} m_{\rm p}c^2\sim150 \ \Gamma_{{\rm rel},0.5}\Gamma_{2.7}\tau_{\rm pn}\ \rm GeV$ in the observer frame~\cite{Murase_et_al_2013}. 
Such quasithermal neutrinos may be detectable by IceCube$+$DeepCore.  

Our goal is to show that a fraction of the injected neutrons recross the shock from the downstream to the upstream, and are accelerated up by the NPC acceleration mechanism. 
The conversion channel is dominated by hadronuclear collisions $p+p\rightarrow n+p+N\pi$ and $n+p\rightarrow p+p+N\pi$ (see~\cite{Derishev_et_al_2003} for other cases).  
Hereafter, we simply assume that (i) a conversion of a nucleon into either of a proton or a neutron occurs with 50\% per each collision, 
(ii) the collision is isotropic in the center-of-mass frame of incident and target nucleons, 
and (iii) the inelasticity and the cross section are independent of the energy, $\kappa_{\rm pn}=0.5$ and $\sigma_{\rm pn}=3\times10^{-26}\ \rm cm^2$, respectively. 

Before proceeding, we should note that our setup includes situations where the the conventional Fermi shock acceleration would be inefficient.  
For $\tau_{\rm T}\gg1$ (note $\tau_T>\tau_{\rm pn}$), the shock may be radiation mediated~\cite{Budnik_et_al_2010, Nakar_Sari_2012}, 
where the shock width or the deceleration length of incoming protons is typically much longer than the isotropization length~\cite{Levinson_Bromberg_2008,Murase_et_al_2011,Murase_Ioka_2013}. 
Such protons cannot perceive the enough jump in the flow velocity, which is crucial for the energy gain.  
On the other hand, the mean free path of elastic and inelastic collisions must be longer than the deceleration length, 
so neutrons are directly injected into the downstream with an initial Lorentz factor $\gamma_{\rm d,o}\sim\Gamma_{\rm rel}$. 
Here, we only consider neutron injections, which gives a conservative estimate on the acceleration efficiency.  

Now let us consider a possible acceleration cycle after the neutron injection.  
From Eq.(\ref{eq:ene_gain}), the energy gain per cycle may be maximized by including the proton phase both in the upstream and the downstream to be isotropized in the magnetic field, 
considering the smallest number of inelastic collisions.  
The optimal cycle is shown in Fig. \ref{f0} (hereafter NPC cycle).
The NPC cycle starts from conversions of injected neutrons into protons in the downstream. 
After being isotropized in the magnetic field, these protons are re-converted to neutrons while they are advected. 
A fraction of the neutrons can cross the shock to the upstream and again can be converted to protons.  
These protons in the upstream are easily captured by the shock, and return back to the downstream.   
Note here that once the protons become nonthermal, the deceleration within the shock width can be neglected when such relativistic protons and electrons are collisionless.  
The energy gain per NPC cycle is $\langle E_{\rm f}/E_{\rm i}\rangle\approx0.5\Gamma_{\rm rel}{}^2$.   

The return probability $P_{\rm ret}$ in the NPC cycle can be roughly estimated as below. 
First, both of the two inelastic collisions must have conversions, which occur with a $1/4$ chance.  Second, only downstream neutrons with $\mu_{\rm d}>\beta_{\rm sh, d}/\beta_{\rm d}$ can cross the shock to the upstream. 
Here $\beta_{\rm sh, d}$ is the shock velocity in the downstream rest frame, which becomes $\approx 1/3$ in the relativistic limit. 
Finally, the fraction of neutrons that experience an inelastic collision in the upstream is $\approx\min[1, \tau_{\rm pn}]$. 
Note that the fraction of protons that leave the upstream is quite small for relatively ordered magnetic fields that we here consider. 
The above arguments yield 
\begin{equation}\label{eq:ret_pro}
P_{\rm ret} \approx f_{\rm npc}\times \frac{1}{12} \min[1, \tau_{\rm pn}]. 
\end{equation}
Here, $f_{\rm npc}$ is a factor to be determined by numerical calculations, including all other uncertainties, 
e.g., the fraction of upstream protons which experience inelastic collisions before being captured by the shock. 

We can define the efficiency of the NPC acceleration as the energies of accelerated nucleons over that of injected neutrons, 
which is given by $\varepsilon_{\rm npc}\approx\kappa_{\rm pn}\times(1/2)\min[1, \tau_{\rm pn}]\times\sum(\langle E_{\rm f}/E_{\rm i}\rangle\times P_{\rm ret})^{N_{\rm cyc}}$. 
Here $N_{\rm cyc}$ is the cycle number, and the pre-factor corresponds the energy loss and the survival fraction at the first conversion. 
As we discuss later, $N_{\rm cyc}$ would be at most a few, considering the Bethe-Heitler~(BH) processes. 
Accordingly, we take only the $N_{\rm cyc} = 1$ component~\cite{Murase_et_al_2013}
\footnote{Definitions in~\cite{Murase_et_al_2013} are somewhat different from those given here. 
The relations between the definitions are given by $\varepsilon_{\rm npc}=(1/4){\rm min}[1,\tau_{\rm pn}]\epsilon_{\rm npc}$ and $f_{\rm npc}=3g_{\rm npc}$.}; 
\begin{equation}\label{eq:eff}
\varepsilon_{\rm npc} \approx f_{\rm npc} \times \frac{\Gamma_{\rm rel}{}^2}{96} \min[1,\tau_{\rm pn}{}^2]. 
\end{equation}
Since the intrinsic energy budget of the accelerated nucleons is the kinetic energy of the
proton flow (rather than that of the neutron flow), $\varepsilon_{\rm npc}$ can become even 
larger than unity, especially for a larger $\Gamma_{\rm rel}$. 

{\it Monte-Carlo simulations.---}
Here we perform Monte-Carlo simulations of the NPC acceleration to justify the estimates above and obtain the energy spectra that depend on details of scattering processes.   

For demonstration, we assume ordered magnetic fields parallel to a plain shock both in the upstream and the downstream, and the compression ratio is the same as the baryon density: 
$B_{\rm d}/B_{\rm u}=n_{\rm d}/n_{\rm u}=4(\Gamma_{\rm rel} +3)$.  
Note that this is not a critical assumption since magnetic fields are relevant just to isotropize protons. 
The downstream temperature can be estimated as  
$T_{\rm d}\approx(n_{\rm u}m_{\rm p}c^2\Gamma_{\rm rel}^2 /a)^{1/4}\sim1\ L_{\rm iso, 52}{}^{1/4}r_{\rm o, 7}{}^{-1/2}\Gamma_{\rm rel, 0.5}{}^{3/2}\Gamma_{2.7}{}^{-1}\ \rm keV$
where $a$ is the radiation constant. 
Consequently, the system is parameterized by $\Gamma_{\rm rel}$, $\tau_{\rm pn}$, and $\xi(1)$. 
Here $\xi(\gamma)\equiv\omega_{\rm g,d}t_{\rm coll,d}=\omega_{\rm g,u}t_{\rm coll,u}$, 
and $\omega_{\rm g}=2\pi eB/\gamma m_{\rm p}c^2$ is the proton-gyration frequency and $t_{\rm coll}{}^{-1}=n\sigma_{\rm pn}c$ is the inelastic-collision frequency.  
When $\xi(\gamma)\gg1$, protons are isotropized before the next inelastic collision. 
In the following calculations, we fix $\xi(1)=10^6$, which corresponds to a conservative magnetic-field strength of $B_{\rm u}\sim4\times10^2\ L_{\rm iso, 51}r_{11.3}{}^{-2}\Gamma_{\rm s, 2}{}^{-2}\ \rm G$. 

We inject $10^7$ neutrons setting the initial Lorentz factor and pitch angle as $\gamma_{\rm d, o}=\Gamma_{\rm rel}$ and $\mu_{\rm d, o}= -1$, respectively,  
and trace the trajectories until the shock sweeps the optical depth $\tau_{\rm pn}$, which corresponds to the dynamical time of the outflow. 
In this case, the adiabatic expansion of the flow would not essentially change our results.

Fig.\ref{f1} shows the energy spectra of protons in the downstream normalized by the neutron injection for a fixed optical depth, $\tau_{\rm pn}=2$.  
The left and right panel shows the case of $\Gamma_{\rm rel}=3$ and $5$, respectively. 
The various bumps correspond to the cycle number $N_{\rm cyc}=0,1,2,3$. 
The $N_{\rm cyc}=0$ peak is at $\approx0.5(\gamma_{\rm d,o}-1)+1$.  
We confirm that the NPC cycle (Fig. \ref{f0}) gives a dominant contribution for the $N_{\rm cyc}\geq1$ components. 
As expected, the peaks are boosted by $\approx0.5\Gamma_{\rm rel}{}^2$ per cycle. 
We also note that the asymptotic energy spectrum becomes harder with a larger $\Gamma_{\rm rel}$, which is different from those predicted by the conventional Fermi acceleration.  
This is essentially due to a larger energy gain per cycle for a larger $\Gamma_{\rm rel}$. 

\begin{figure}
\includegraphics[width=3.25in]{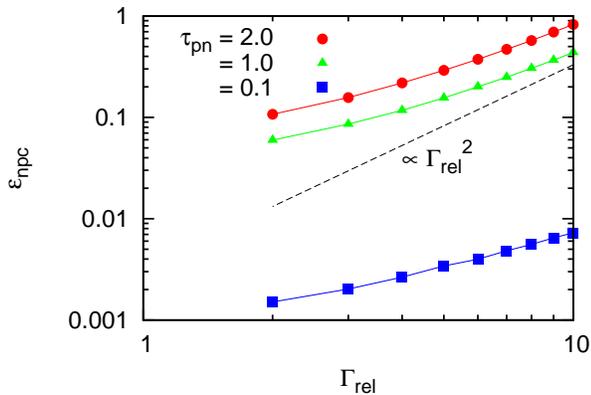}
\caption{The efficiency of the NPC acceleration. 
The total energy of accelerated baryons by a single cycle is normalized by that of the neutron injection.  
We fix $\gamma_{\rm d,o}=\Gamma_{\rm rel}$ and $\xi(1)=10^6$.
The circles, triangles, and squares correspond to  $\tau_{\rm pn}=0.1,1,{\rm and}\ 2$, respectively.}  
\label{f2}
\vspace{-1.\baselineskip}
\end{figure}

Fig. \ref{f2} shows the acceleration efficiency, where we only take the $N_{\rm cyc}=1$ component.  The circles, triangles, and squares correspond to $\tau_{\rm pn}=0.1,1,{\rm and}\ 2$, respectively.  
Note that $\varepsilon_{\rm npc}\propto\tau_{\rm pn}{}^2$ for a fixed $\Gamma_{\rm rel}$, 
and $\varepsilon_{\rm npc}\propto\Gamma_{\rm rel}{}^{2}$ for a fixed $\tau_{\rm pn}$ especially at a larger $\Gamma_{\rm rel}$, which is consistent with Eq. (\ref{eq:eff}) with $f_{\rm npc}\sim0.1\mbox{-}1$. 
One can see the enhancement of $\varepsilon_{\rm npc}$ at lower $\Gamma_{\rm rel}$. 
We find that this comes from a different path from the NPC cycle, in which injected neutrons cross the shock form the down stream to the upstream by experiencing a large-angle scattering.      
Thanks to the smaller number of the inelastic collision, the energy gain in the above path can be a factor $2$ larger than the NPC cycle.  
Such components, however, become smaller for a larger $\Gamma_{\rm rel}$ where most of the scattered neutrons are still directed to the far downstream.

{\it Summary and discussion.---}
We numerically investigated the NPC acceleration mechanism. 
It may be relevant at internal shocks occurring in neutron-loaded relativistic outflow even in the radiation-mediated regime, where the conventional Fermi shock acceleration would be inefficient~\cite{Levinson_Bromberg_2008,Murase_et_al_2011,Murase_Ioka_2013}.  
We showed that $\sim\Gamma_{\rm rel}{}^2\min[1,\tau_{\rm pn}{}^2]$\% of the neutron-flow energy may be converted to non-thermal nucleons with boosts of $\gtrsim0.5\Gamma_{\rm rel}{}^2$.  

So far, we only took into account the hadronuclear collision.  In fact, other energy-loss processes may determine the maximum energy obtained by the NPC acceleration.  
In the case of GRBs, the BH process $p+\gamma\rightarrow p+e^{-}+e^{+}$ would become crucial for sufficiently high-energy protons. 
For a black-body spectrum, this gives a maximum Lorentz factor of $\gamma_{\rm d,max}\lesssim 2m_{\rm e}c^2/Ck_{\rm B}T_{\rm d}$, 
where $C$ is the pre-factor taking into account the effect of the Wien tail. 
In addition, the NPC acceleration becomes inefficient for $\xi(\gamma_{\rm u(d)})\lesssim1$, 
where the pitch angle of a proton is no longer isotropized before the next conversion or crossing the shock.  
Then, it becomes difficult to cross the shock from the downstream to the upstream. 
Also, the typical pitch angle in the upstream becomes $\langle\mu_{\rm u}\rangle\approx1-1/\Gamma_{\rm rel}^2$ as in the case of the Fermi acceleration, 
which makes the energy gain per cycle negative, $\langle E_{\rm f}/E_{\rm i}\rangle<1$, due to the inelasticity of the collisions.  
This sets another constraint of $\gamma_{\rm d,max}\lesssim\xi(1)$.  Consequently, the maximum Lorentz factor by the NPC acceleration can be described as 
\begin{equation}
\gamma_{\rm d,max} \approx \min \left[ \frac{2m_{\rm e} c^2}{C k_{\rm B} T_{\rm d}}, \frac{eB_{\rm u}}{\sigma_{\rm pn} m_{\rm p} c^2 {n_{\rm u}}}  \right].
\end{equation}
For instance, substituting $\Gamma=600$, $\Gamma_{\rm rel}=3$, $\tau_{\rm pn}=1$, and $\xi(1)=10^6$, 
which is a possible parameter set for a {\it successful} GRB jet~\cite{Murase_et_al_2013}, the NPC acceleration can give $\gamma_{\rm d,max}\sim 200$ if $C\sim6$.  
The by-product neutrino energy can be $E^{\rm obs}_{\nu}\approx0.05\Gamma\gamma_{\rm d}m_{\rm p}c^2\sim 2 \ \Gamma_{2.7}\gamma_{\rm d,2.3}~\rm TeV$ in the observer frame. 
Such a high-energy tail is crucial for the detection of subphotospheric neutrinos from GRBs as shown in \citep{Murase_et_al_2013}.  

In this work, we adopted a test-particle approximation assuming that the neutron fraction is less than unity, where the backreaction on the background shock structure is neglected.  
Once the total energy or pressure of accelerated nucleons becomes significant compared to that of the proton flow (rather than the neutron flow), 
inelastic collisions in the upstream contribute to deceleration of the proton flow with the length scale $\approx1/n_{\rm u}\sigma_{\rm pn}$ and the results should be affected. 

Also, we assumed ordered magnetic fields for the Monte-Carlo simulations.  
One can expect turbulent magnetic fields especially in the shock downstream where the proton diffusion has to be considered. 
We note that our results would not change much if the diffusion velocity is so slow that the protons cannot cross the shock to the upstream. 
If not, the conventional shock acceleration can work effectively after the neutron injection. 
Those cases will be investigated in future work. 

In addition, we treated the inelastic interactions based on the simplified assumptions (i)-(iii).  
Assumption (i) is not strictly valid in lower energies, 
where the conversion processes from protons to neutrons occur slightly less frequently in total than in nonconversion processes~\citep{1986A&A...157..223D}. 
By this effect, the efficiency of the NPC acceleration can be affected slightly. 
As for assumption (ii), the scatterings occur in an anisotropic manner even in the center-of-mass frame,  
and the resultant energy spectra can become more peaky because of the directionality. 
However, for $\xi(1) \gg 1$, the effect of such anisotropic scatterings can be smeared out by the gyration in the proton phase.
Assumption (iii) becomes invalid in higher energies~\cite{1986A&A...157..223D,1981Ap&SS..76..213S}.  
However, as argued above, the NPC acceleration is practically effective only below $\gamma_{\rm d, max}$, 
where assumption (iii) is typically a good approximation. 

Finally, we should remark that the NPC acceleration may operate in {\it failed} GRB jets~\cite{2001PhRvL..87q1102M}, protoneutron star winds~\cite{2013arXiv1303.2612M} buried in the progenitor, 
and possibly active galactic nuclei jets. 

{\it Acknowledgments.---}
This work is supported by a JSPS fellowship for research abroad~(KK), NASA NNX13AH50G~(PM) and 
the Hubble Fellowship grant No. 51310.01 awarded by the Space Telescope Science Institute, 
which is operated by the Association of Universities for Research in Astronomy, Inc., for NASA, under contract NAS 5-26555~(KM).  
We acknowledge the support by CCAPP workshop, Revealing Deaths of Massive Stars with GeV-TeV Neutrinos.


\end{document}